\documentclass[runningheads]{llncs}
\usepackage{graphicx}
\begin{document}
\title{B-line Detection in Lung Ultrasound Videos: Cartesian vs Polar Representation}
%

\author{Hamideh Kerdegari\inst{1} \and
Phung Tran Huy Nhat\inst{1,2} \and
Angela McBride\inst{2} \and
Luigi Pisani\inst{3} \and
Reza Razavi\inst{1} \and
Louise Thwaites\inst{2} \and
Sophie Yacoub\inst{2} \and
Alberto Gomez\inst{1} }
\authorrunning{H. Kerdegari et al.}
%
\institute{School of Biomedical Engineering \& Imaging Sciences, King's College London, UK \and
 Oxford University Clinical Research Unit, Ho Chi Minh City, Vietnam \and
 Mahidol Oxford Research Unit, Thailand}


\maketitle              
\begin{abstract}
Lung ultrasound (LUS) imaging is becoming popular in the intensive care units (ICU) for assessing lung abnormalities such as the appearance of B-line artefacts as a result of severe dengue. These artefacts appear in the LUS images and disappear quickly, making their manual detection very challenging. They also extend radially following the propagation of the sound waves. As a result, we hypothesize that a polar representation may be more adequate for automatic image analysis of these images. This paper presents an attention-based Convolutional+LSTM model to automatically detect B-lines in LUS videos, comparing performance when image data is taken in Cartesian and polar representations. Results indicate that the proposed framework with polar representation achieves competitive performance compared to the Cartesian representation for B-line classification and that attention mechanism can provide better localization.

\end{abstract}
\section{Introduction}

\begin{figure}
\centering
\includegraphics[width=\linewidth]{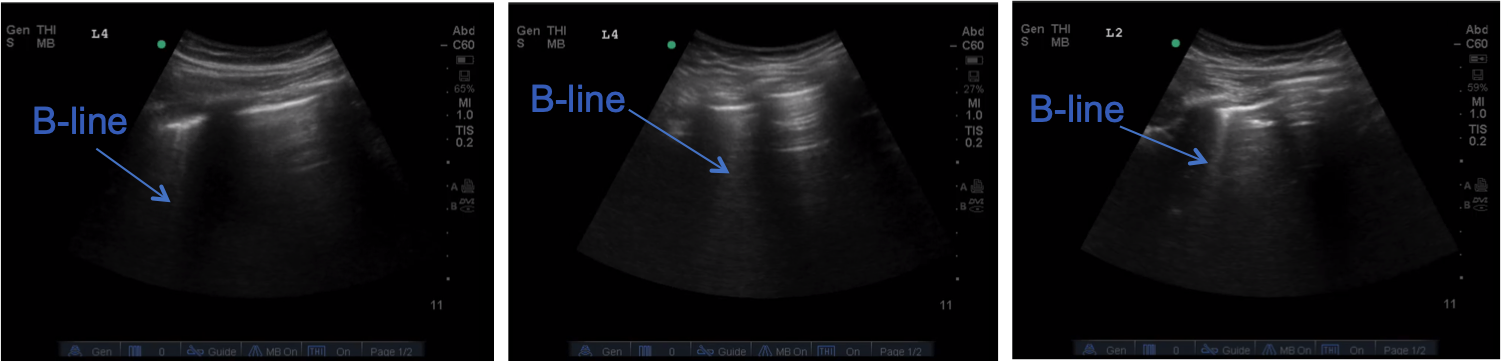}
\centering
\caption{Examples of LUS B-line frames. B-line artefacts are presented as bright lines that develop from the surface of the lung.} 
\label{3exa}
\end{figure}

Recently, lung ultrasound (LUS) imaging has increased in popularity  for rapid lung monitoring in patients in the intensive care units (ICU). Particularly for dengue patients, LUS can capture image artefacts such as B-lines that indicate a pulmonary abnormalities such as oedema and effusions \cite{soldati2019ultrasound}. B-lines are bright lines extending from the surface of the lung distally following the direction of propagation of the sound wave (shown in Figure \ref{3exa}). LUS imaging is useful for assessing lung abnormalities though the presence of B-lines. However, these lines become visible randomly during respiratory cycle in the affected area only \cite{dietrich2016lung}; therefore, manually detecting these artefacts becomes challenging for inexperienced sonographers, and particularly in low and middle income countries with higher prevalence of these diseases where training opportunities and experise are scarce.

In order to provide an automatic solution to the LUS B-line detection problem, recent studies proposed classification, segmentation and localization of B-line artefacts in individual LUS frames. For example, a convolutional neural network (CNN) followed by a class activation map was proposed by Sloun et al. \cite{van2019localizing} to classify B-lines and produce a segmentation map of them, respectively. A weakly supervised localization of B-lines was proposed using a spatial transformer network \cite{roy2020deep}. In another study \cite{kulhare2018ultrasound}, a single-shot CNN was used to localize B-lines with bounding boxes. Previous work by Kerdegari et al. \cite{Hamideh2021-Bline} showed that employing temporal information can improve B-line detection task in LUS, leveraging a temporal attention mechanism to localize B-line frames within LUS videos.

Furthermore, attention mechanisms have been used widely for spatial localization of lung lesions particularly in CT and x-ray lung images. For instance, a residual attention U-Net for multi-class segmentation of CT images \cite{chen2020residual} and x-ray images \cite{gaal2020attention} was proposed. A lesion-attention deep neural network (LA-DNN) was presented by \cite{liu2020online} to do two tasks of B-line classification and multi-label attention localization of five lesions. All these studies employed spatial attention mechanism for lung lesion localization. 

LUS images are usually used in a standard Cartesian coordinate representation (i.e., scan-converted). In this representation, the B-lines commonly appear densely in the middle of frustum. Therefore, data preprocessing techniques such as downsampling might cause information loss with Cartesian representation. Additionally, the radial direction that B-lines follow is known but this information is not exploited. In this paper, we propose to use a polar representation to, first, reduce information loss when downsampling the data, and second, leverage prior knowledge about line formation by having one dimension aligned with the lines.

To this end, we compare the performance of the temporal attention-based convolutional+LSTM model proposed by \cite{Hamideh2021-Bline} when using Cartesian and polar representations. In summary, the contribution of this paper is investigating the effect of using LUS polar coordinate representation on the B-line detection and localization performance. Also, we evaluate the effect of different downsampling factors of LUS video with polar and Cartesian representations for B-line detection and localization tasks.

\section{Model Architecture}

\begin{figure}
\centering
\includegraphics[width=0.9\linewidth]{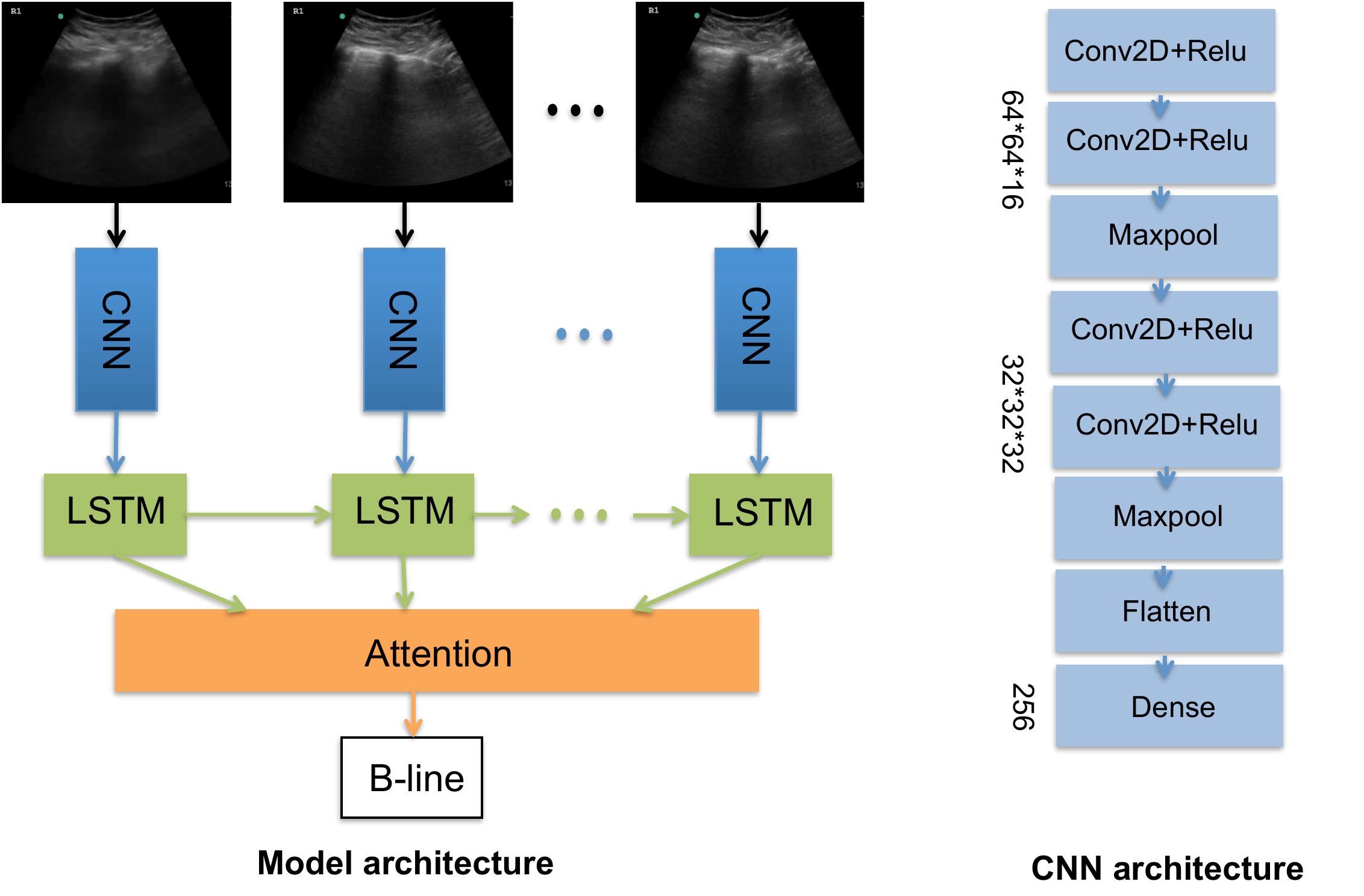}
\centering
\caption{Overview of the proposed framework, Left: LUS videos are first processed through CNN layers, then a bidirectional LSTM is used to extract temporal information and finally an attention mechanism is applied to localize B-line frames within LUS videos. Right: Detailed architecture of our CNN.} 
\label{modcnn}
\end{figure}

This paper employs a model that combines a deep visual feature extractor such as a CNN with a long short-term memory (LSTM) network that can learn to recognize temporal dynamics of videos; and a temporal attention mechanism to learn where to pay more attention in the video. Figure \ref{modcnn} shows the core of our model. This model works by passing each frame from the video through our CNN model (The architecture details are explained in Figure \ref{modcnn}, right) to produce a fixed length feature vector representation. The outputs of our CNN are passed into a bidirectional LSTM (16 hidden units, tanh activation function) network as a recurrent sequence learning model. Then, the LSTM outputs are passed to the attention network \cite{bahdanau2014neural} to produce an attention score ($e_{t}$) for each attended frame ($h_{t}$): \(e_t=h_t w_a\), where $w_{a}$ represents attention layer weight matrix. From $e_{t}$, an importance attention weight ($a_{t}$) is computed for each attended frame: \(a_t = \frac{\exp (e_{t})}{\sum_{i=1}^{T}\exp (e_{i})}\). To learn which frame of the video to pay attention to, $a_{t}$s are multiplied with the LSTM output. Finally, the output of LUS video classification is generated by averaging the attention weighted temporal feature vector over time and passing to a fully connected layer for video classification.

\section{Experimental Setup}

\subsection{Dataset and Preprocessing}
The dataset used in the experiments was collected at the Hospital of Tropical Diseases in Ho Chi Minh City, Vietnam. It includes about 5 hours of lung ultrasound videos collected from 60 dengue patients. These videos were collected using a Sonosite M-Turbo machine (Fujifilm Sonosite, Inc., Bothell, WA) with a low-medium frequency (3.5-5 MHz) convex probe. The Kigali ARDS protocol \cite{riviello2016hospital}, as a standardised operating procedure was applied at 6 points (2 anterior, 2 lateral and 2 posterolateral) on each side of the chest to perform LUS exams.

The four-second LUS video clips have been resized from original size of $640 \times 480$ pixels to $64 \times 64$ pixels for training, and fully anonymised through masking. A qualified sonographer annotated these clips using the VGG annotator tool \cite{dutta2019vgg}. During the annotation procedure, each video clip was annotated by being assigned either a B-line or non-B-line label. Further, B-line frames and B-line regions in the B-line video clips were annotated to be used as annotations for temporal and spatial B-line localization task later. 

\subsection{Polar Coordinate Representation}
\begin{figure}
\includegraphics[width=0.9\linewidth]{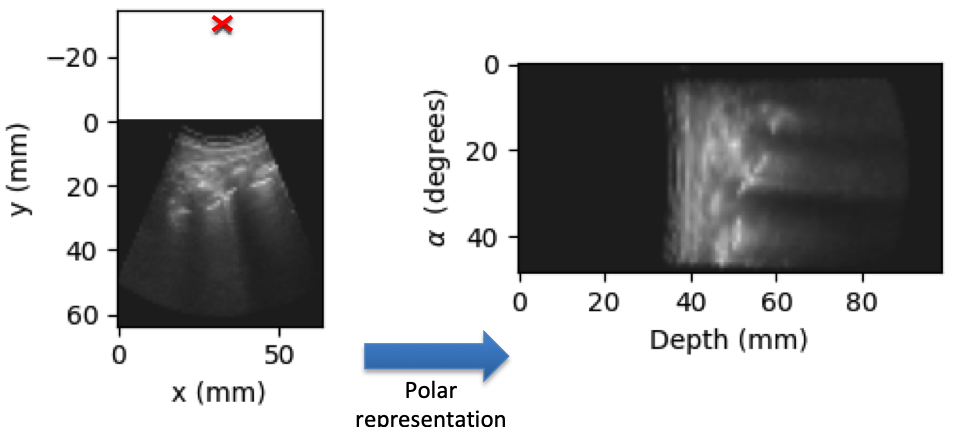}
\centering
\caption{Examples of a B-line frame in Cartesian coordinate (left) and polar coordinate (right) representation.} 
\label{PolarRep}
\end{figure}

Like other common applications of ultrasound imaging, lung ultrasound images are normally presented in Cartesian coordinates (shown in Figure \ref{PolarRep}, left). In this case, the information particularly B-line artefacts are presented densely in the centre of the frustum to some extend. Therefore, when we downsample the LUS videos as input to our network some information are lost. To overcome this limitation, we transform each video clip into its associated polar coordinate representation (shown in Figure \ref{PolarRep}, right). With polar coordinate representation, information are expanded along the degree axes of polar data; therefore, less information are missed during downsampling of the data. Additionally, there is not much information in the left and right up corner (black areas) of Cartiesian coordinate representation. As a result, when these areas are removed in the polar coordinate representation, the network can concentrate on the areas of each frame where more useful information are exist.

Polar representation is carried out by the following reparameterization, used to resample the Cartesian images into a polar grid using bilinear interpolation:
\begin{equation}
\begin{array}{rl}
     x=& r \sin(\alpha) \\
     y=& r \cos(\alpha) 
\end{array}
\end{equation}
Where $r$ is the depth, or radius (distance form the beam source to a pixel location) and $\alpha$ is the angle measured from the y axis. 
\subsection{Implementation Details}
Our network was implemented using Keras library with a Tensorflow backend. The network was optimised using Adam optimizer with the learning rate (lr) set to $10^{-6}$. Batch normalization was used for both CNN and LSTM parts of the network. Batch size of 25, dropout of 0.2 and $L2 = 10^{-5}$ for regularization were employed. Data augmentation was applied to the training data by adding horizontally-flipped frames. 5-fold cross validation was used and the network converged after 60 epochs. The class imbalance was addressed by weighting the probability to draw a sample by its relative class occurrence in the training set.

\section{Experiments and Results}
In order to investigate the potential benefit of employing polar representation and various video resolutions in B-line detection task, we trained our model with Cartesian and polar representations using various input video sizes of $64\times 64$, $32\times 32$ and $16\times 16$ resolution. Furthermore, we reduced the depth size of polar data to 32 and 16 samples, while keeping the number of angular elements to 64 (hence maintaining angle resolution), therefore having the video size of $64\times 32$ and $64\times 16$ resolution for training.

\begin{figure}
\includegraphics[width=0.9\linewidth]{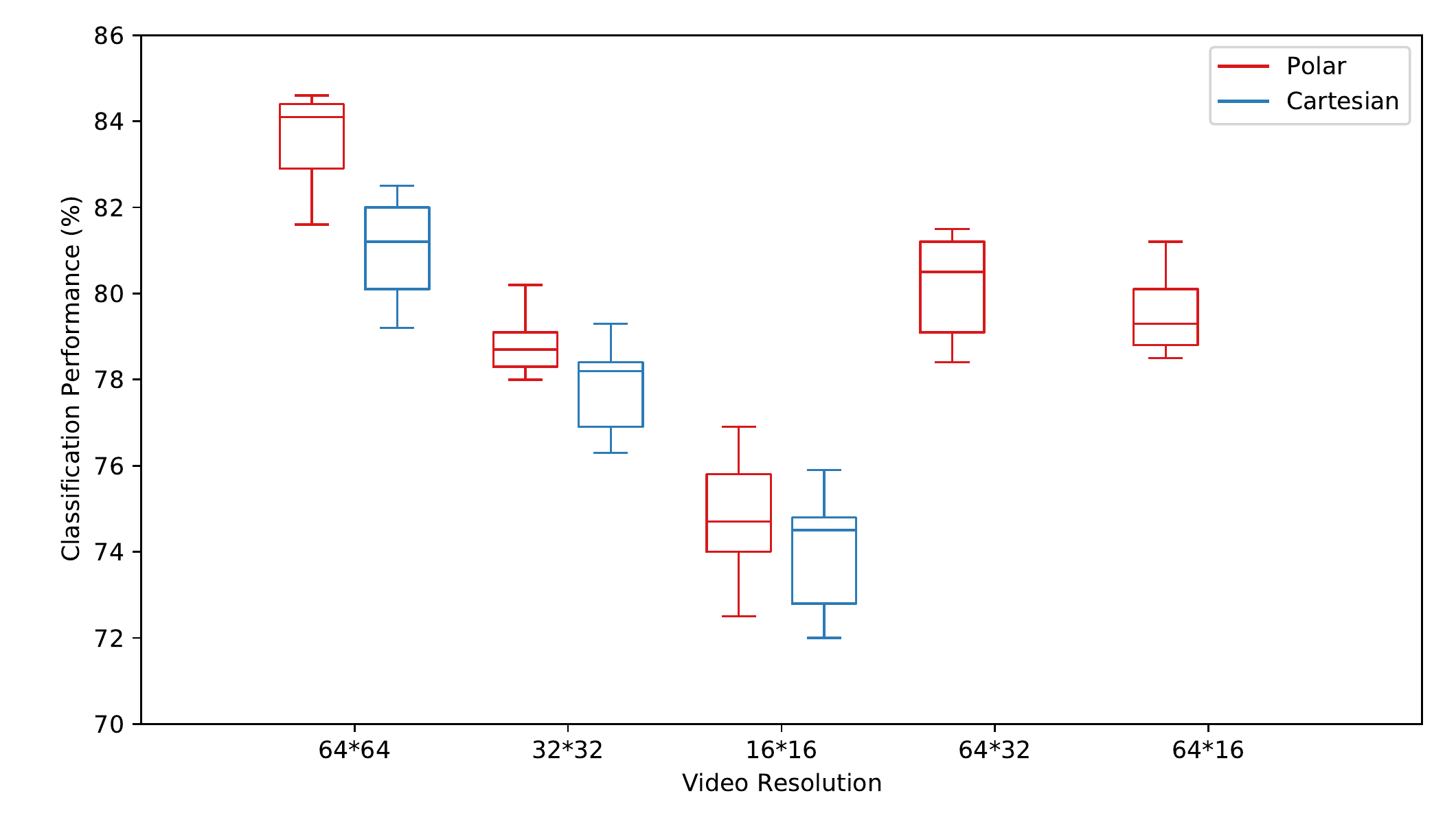}
\centering
\caption{B-line classification performance (F1 score) of Cartesian and polar representations with various video resolutions.} 
\label{perf}
\end{figure}

To assess the classification performance of the model, the harmonic mean of precision and recall expressed as $F1= 2\times \frac{Precision\times Recall}{Precision + Recall } \times 100$ score (\%) was utilised.
The classification performance for Cartesian and polar data are presented in Figure \ref{perf}. An alpha value of 0.05 was selected as the statistical significance threshold. Shapiro-Wilk test showed that all data were normally distributed.
Our baseline video resolution ($64\times 64$) received the highest performance for both polar and Cartesian representations. Also, a paired t-test revealed that the performance of polar data (83.5\%) is significantly higher than Cartesian data (81\%) (t=2.776, p=0.017) in all cases with the same number of pixels. This demonstrates that the model can extract more information from a polar representation. When we decreased the video resolution into $32\times 32$ and $16\times 16$, the performance dropped compared to the baseline video resolution, although the drop was less significant in polar images. For video resolutions of $32\times 32$, paired t-test showed significant difference between Cartesian and polar representation in B-line detection task (t=1.035 , p=0.028). However, this difference is not significant for video resolution of $16\times 16$ (t=-1.104, p=0.165), probably because the downsampling is too aggressive and B-lines become barely distinguishable in any representation. Furthermore, we decreased the depth size of polar data into 32 and 16 to evaluate the contribution of depth information in B-line detection. Compared to the depth size of 64 in baseline resolution, the performance decreased significantly for both depth sizes of 32 (t=2.835 , p=0.008) and 16 (t=1.503 , p=0.018). 

Additionally, we investigated the impact of downsampling along scan-lines and along angles. To do this, we compared two video resolutions that had the same amount of pixels: $32\times 32$ and $64\times 16$. Results showed that video resolution of $64\times 16$ (64 along the angle dimension) has significantly higher performance which shows that preserving information along the angle dimension helps in this specific task where artefacts are aligned along constant-angle lines (t=2.43, p=0.03).

We further evaluated B-line temporal localization accuracy using both data representations. We calculated intersection over union (IoU) of predicted temporal localized frames with their ground truth annotations. Results are presented in Table \ref{tab:IOU}. With polar representation, the model is able to localize B-line frames temporally with higher performance than Cartesian representation. Additionally, the attention weights for true B-line frames are higher in polar representation and for the non B-line frames lower, compared to Cartesian representation, further suggesting that the network learns to differentiate B-line and non B-line frames better in polar representation.

\begin{table}[]
\normalsize
\caption{IoU values showing B-line localisation accuracy (\%) for various video resolutions of Cartesian and polar representations.}
\label{tab:IOU}
\centering
\addtolength{\tabcolsep}{7pt} 
\begin{tabular}{llllll}
\hline
\multicolumn{6}{c}{\textbf{Video Resolution}}                                                                             \\ \hline
                   & \textit{64*64} & \textit{32*32} & \textit{16*16} & 64*32                   & 64*16                   \\ \hline
\textbf{Cartesian} & 67.1           & 56.3           & 42.2           & \multicolumn{1}{c}{---} & \multicolumn{1}{c}{---} \\ \hline
\textbf{Polar}     & 73.2           & 62.5           & 43.1           & 67.7                    & 65.1                    \\ \hline
\end{tabular}
\addtolength{\tabcolsep}{6pt}
\end{table}


\section{Conclusion}

This paper investigates the effect of employing ultrasound polar coordinate representation on LUS B-line detection and localization tasks. We employed an attention-based convloutional+LSTM model capable of extracting spatial and temporal features from LUS videos and localizing B-line frames using a temporal attention mechanism. We evaluated B-line classification and localization with this architecture using Cartesian and polar coordinate representations with different resolutions. Using our LUS video dataset, results showed that polar representation consistently outperforms Cartesian in terms of classification accuracy and temporal localization accuracy. 

Our future work will explore an spatiotemporal attention mechanism that is able to detect the B-line artefacts and localize them both spatially and temporally within LUS videos in polar coordinates. B-line spatial localization may help clinicians to quantify the severity of the disease. Overall, these findings will assist management of ICU patients with dengue particularly in low- and middle-income countries where ultrasound operator expertise is limited.

\section*{ACKNOWLEDGMENT}
The VITAL Consortium: \textbf{OUCRU}: Dang Trung Kien, Dong Huu Khanh Trinh, Joseph Donovan, Du Hong Duc, Ronald Geskus, Ho Bich Hai, Ho Quang Chanh, Ho Van Hien, Hoang Minh Tu Van, Huynh Trung Trieu, Evelyne Kestelyn, Lam Minh Yen, Le Nguyen Thanh Nhan, Le Thanh Phuong, Luu Phuoc An, Nguyen Lam Vuong, Nguyen Than Ha Quyen, Nguyen Thanh Ngoc, Nguyen Thi Le Thanh, Nguyen Thi Phuong Dung, Ninh Thi Thanh Van, Pham Thi Lieu, Phan Nguyen Quoc Khanh, Phung Khanh Lam, Phung Tran Huy Nhat, Guy Thwaites, Louise Thwaites, Tran Minh Duc, Trinh Manh Hung, Hugo Turner, Jennifer Ilo Van Nuil, Sophie Yacoub. \textbf{Hospital for Tropical Diseases, Ho Chi Minh City}: Cao Thi Tam, Duong Bich Thuy, Ha Thi Hai Duong, Ho Dang Trung Nghia, Le Buu Chau, Le Ngoc Minh Thu, Le Thi Mai Thao, Luong Thi Hue Tai, Nguyen Hoan Phu, Nguyen Quoc Viet, Nguyen Thanh Nguyen, Nguyen Thanh Phong, Nguyen Thi Kim Anh, Nguyen Van Hao, Nguyen Van Thanh Duoc, Nguyen Van Vinh Chau, Pham Kieu Nguyet Oanh, Phan Tu Qui, Phan Vinh Tho, Truong Thi Phuong Thao. \textbf{University of Oxford}: David Clifton, Mike English, Heloise Greeff, Huiqi Lu, Jacob McKnight, Chris Paton. \textbf{Imperial College London}: Pantellis Georgiou, Bernard Hernandez Perez, Kerri Hill-Cawthorne, Alison Holmes, Stefan Karolcik, Damien Ming, Nicolas Moser, Jesus Rodriguez Manzano. \textbf{King’s College London}: Alberto Gomez, Hamideh Kerdegari, Marc Modat, Reza Razavi. \textbf{ETH Zurich}: Abhilash Guru Dutt, Walter Karlen, Michaela Verling, Elias Wicki. \textbf{Melbourne University}: Linda Denehy, Thomas Rollinson.

\end{document}